\documentclass[runningheads]{llncs}
\usepackage[T1]{fontenc}
\usepackage{lmodern}
\usepackage{amssymb}
\usepackage{microtype}
\usepackage{booktabs}
\usepackage{xurl}
\usepackage{xcolor}
\definecolor{linkblue}{RGB}{26,13,171}
\usepackage[bookmarks=false,colorlinks,linkcolor=linkblue,citecolor=linkblue,urlcolor=linkblue,
  pdftitle={Short Paper: Prefix Count Limits Can Increase First-Hit Discovery in Card Reissuance},
  pdfauthor={Wasif Faisal and Suprava Saha Dibya},
  pdfsubject={payment-card security, enumeration, card reissuance},
  pdfkeywords={payment-card security, payment account enumeration, card reissuance, security metrics}]{hyperref}

\newcommand{\CostExtraMean}{596.03}
\newcommand{\AdditionsMeanFour}{200.17}
\newcommand{\MatchedReduction}{1.75}
\newcommand{\MatchedLow}{1.62}
\newcommand{\MatchedHigh}{1.88}
\newcommand{\MechanismMeanBefore}{82.38}
\newcommand{\MechanismMaxBefore}{99.42}
\newcommand{\MechanismAfter}{94.47}

\begin{document}
\title{Short Paper: Prefix Count Limits Can Increase First-Hit Discovery in Card Reissuance}
\titlerunning{Count-Limiting and Discovery}
\author{Wasif Faisal \and Suprava Saha Dibya}
\authorrunning{W. Faisal and S. S. Dibya}
\institute{BRAC University\\
\email{md.wasif.faisal@g.bracu.ac.bd}\\
\email{ext.suprava.saha@bracu.ac.bd}}
\maketitle

\begin{abstract}
  When a card number is compromised, an attacker may search for active numbers sharing its prefix. An issuer might respond by reissuing cards from heavily populated prefixes into less populated ones. We show that this intuitive count control can backfire. For fixed search regions, exposure weights, and total activity, we derive exactly when reducing the maximum prefix count increases the chance that a budgeted search finds an active number. In matched synthetic simulations over 50{,}000 candidates, targeted replacement meets the count limit in every run but raises supplied-12-digit-prefix discovery relative to equal-volume random replacement in three of twelve settings. Thus a lower prefix count does not by itself certify lower enumeration risk.
\end{abstract}

\section{Introduction}
Guessing candidates to identify active card numbers is called \emph{enumeration} \cite{visa_enumeration_2021}. The 2016 Tesco Bank attack makes concentration risk concrete. The FCA found that exhausting randomly populated 50,000-number batches left thousands of valid sequential card numbers and made authentic numbers easier to identify; attackers likely generated card numbers, affected 8,261 accounts, and obtained \pounds2.26 million. Transactions on replaced cards also bypassed fraud analysis because controls operated at account rather than card level \cite[paras.~4.7, 4.16, 4.24]{fca_tesco_2018}. Visa reports enumeration as a leading threat concentrated within BIN ranges; Netcetera describes testing above ten requests per second that doubled a bank's normal traffic and reports that 3-D Secure responses can reveal PAN existence without successful authentication \cite{visa_perc_2025,visa_vaai_2023,netcetera_bin_attack_2025}. Visa guidance combines narrower or consolidated assignment with anti-enumeration advice against sequential PANs \cite{visa_bin_assignment,visa_enumeration_2021}, and an issuer notice documents reissuance into new ranges \cite{austin_fcu_bin_2025}.

Existing guidance addresses card-number assignment and enumeration but does not show that lowering the maximum active count under a prefix lowers attacker success. We use count-targeted redistribution to audit this missing implication, not to claim deployed issuer practice.

A compromised PAN exposes all its leading prefixes after cancellation. We study 11- and 12-digit regions, containing 10,000 and 1,000 Luhn-valid completions. Longer prefixes contain fewer candidates but may also contain fewer active accounts. We ask whether moving cards out of crowded prefixes reduces budgeted first-hit success relative to equal-volume random replacement. Search seeks any active number sharing the supplied original's prefix, not its replacement, under fixed exposure weights. Diminishing returns can make redistribution raise weighted discovery while lowering the maximum count. The result also covers reassignment or migration.

\paragraph{Contributions.}
We derive the exact transfer condition; distinguish feasibility, inventory completion, cap attainment, and discovery; and reproduce reversals in matched equal-volume experiments across shapes, densities, and budgets. Issuers should audit attacker-conditioned discovery rather than treat lower prefix counts as a security certificate.

\paragraph{Related work.}
Prior work models breach-triggered reissuance, distributed guessing of auxiliary fields for known PANs, and virtual cards \cite{graves2018too,ali2017distributed,molloy2007dynamic}. We instead study discovery of another active PAN. CardSim motivates synthetic data, not our allocations \cite{allen2025cardsim}. Guessing work shows that budgeted rankings depend on scenario and strategy; gain functions formalize that dependence \cite{bonneau2012guessing,wang2023silver,alvim2012gain,pendleton2016metrics}. We treat $P_t(q)$ as expected binary first-hit gain for this scenario, not a scenario-independent security score, when auditing balanced allocation \cite{azar1999balanced,bansal2022deletions}.

\section{Explicit Numbers, Replacement, and Search}
\label{sec:model}
\paragraph{The candidate numbers.}
Each 16-digit synthetic candidate has six fixed zeros, a six-digit bucket index, a three-digit suffix, and one Luhn digit ($6+6+3+1=16$), preserving leading zeros. The initial zeros are not a BIN variable, and neither the experiment nor its buckets represent six- or eight-digit BINs or an issuer portfolio. These are not processor-approved test cards. Each bucket has 1,000 candidates sharing a 12-digit prefix; ten adjacent buckets share an 11-digit prefix. Fifty buckets form the experimental space, not a full BIN range. The analysis permits general bucket capacity $M$; 1,000 is only the experimental instantiation. With a 16-digit PAN, $p$ known leading digits leave $10^{15-p}$ Luhn-valid completions; six to eight digits shrink a local region from $10^9$ to $10^7$, but not success if activity falls proportionally \cite{visa_bin_assignment}.

A candidate number is \emph{active} when assigned to an account; all others are inactive. Density is the active fraction, and a valid checksum does not establish activity \cite{pci_luhn_faq}. SAS describes candidate generation from known BIN and expiration information and detection by a shared 12-digit prefix \cite[pp.~3, 8]{sas2021bin}. We use this only as precedent for a local search region; retired-card exposure is our audit scenario, and a BIN alone does not supply the longer prefix.

\paragraph{Replacing a number.}
The \emph{register} holds one active number per account, sampled without repetition within buckets. Replacement reserves a never-assigned number before retiring the old and activating the new. Assignment stays within a bucket, group, or the whole space. Failed reservation stops the campaign without changing the register. Retired numbers remain searchable but cannot be reassigned during the run, a model assumption rather than universal practice.

\paragraph{Searching for active numbers.}
With assignments fixed, uniform search without repetition stops at the first active number or $q$ examinations. It covers the whole space, a known most-crowded bucket, or a supplied prefix (Section~\ref{sec:retired-prefix}). Candidates are unranked and retired numbers unidentified. The counterexample invalidates the count-only implication for this simple attacker; rankings may change policy orderings.

The Netcetera incident supplies operational precedent for high-rate search and recognition. We idealize the recognition channel as error-free, an attacker-favorable benchmark not universal to merchants or ACSs. The model has no authorization process, expiration dates, security codes, bank data, or payment calls; real declines have multiple causes \cite{stripe_decline_codes}. Allocations are scenarios, not issuer estimates. Budgets count candidates, not transactions or time; we neither recover a replacement nor switch regions. Discovery is not itself fraud, but can enable subsequent account testing or fraudulent use \cite{visa_enumeration_2021,ali2017distributed}.

\section{The Audit}
\label{sec:audit}
\paragraph{A balancing example.}
Two prefixes each contain 999 candidates after excluding one retired number. Moving their active counts from $(400,100)$ to $(250,250)$ lowers the larger ten-guess discovery probability from \MechanismMaxBefore{}\% to \MechanismAfter{}\%. But if either prefix is equally likely to be searched, average success rises from \MechanismMeanBefore{}\% to \MechanismAfter{}\%, a 12.09-point increase. Balancing helps one metric and worsens another.

\paragraph{Budgeted first-hit success.}
Region $B$ contains $M(B)$ candidates, with $A_t(B)$ active at time $t$. For integers $M\geq1$, $0\leq a\leq M$, and $0\leq q\leq M$, uniform search without replacement succeeds with probability
\begin{equation}
  F(M,a,q)=1-\frac{{M-a\choose q}}{{M\choose q}},\qquad
  F_t(B,q)=F(M(B),A_t(B),q).
  \label{eq:budget}
\end{equation}
The ratio is the probability of all-inactive draws; ${n\choose q}=0$ for $q>n\geq0$. Discovery is impossible if $a=0$ or $q=0$ and concerns any active number, not a particular replacement.

For partition $\mathcal P$, let $W_t(q)=\max_{B\in\mathcal P}F_t(B,q)$, with $q$ within each region's capacity.

\paragraph{Prefix-weighted discovery.}
Fix $N>0$ original card numbers. Each episode supplies one uniformly, then searches its prefix for at most $q$ candidates, excluding only that number. Bucket $b$ contains $n_b$ original numbers, giving fixed weights $w_b=n_b/N$ across policies and time. After their retirement, for $0\leq q\leq M-1$,
\begin{equation}
  P_t(q)=\sum_{b:\,n_b>0}w_b F(M-1,A_t(b),q).
  \label{eq:prefix-weighted}
\end{equation}
These are not measured attack frequencies. They define one evaluation scenario: exposure inherited from the cancelled-number population, followed by uniform search within the supplied prefix. Fixed weights isolate allocation from exposure selection; policy-dependent exposure needs a joint model. Different prefix-selection strategies or gain functions can rank policies differently. Before retirement, excluding an active supplied number also subtracts one active candidate.

\paragraph{When a balancing transfer increases discovery.}
With weights fixed and the supplied originals retired, consider a feasible transfer between two represented buckets $j$ and $k$, with $a_k<a_j$. Define $\delta^+=F(M-1,a_k+1,q)-F(M-1,a_k,q)$ and $\delta^-=F(M-1,a_j,q)-F(M-1,a_j-1,q)$. Then
\begin{equation}
  \Delta P(q)=w_k\delta^+-w_j\delta^-.
  \label{eq:transfer}
\end{equation}
For $q\geq1$, the marginal increase at count $a$ is ${M-2-a\choose q-1}/{M-1\choose q}$, nonincreasing in $a$, so $\delta^+\geq\delta^-$. Equality is possible for adjacent counts or saturated probabilities; for $q=0$ both marginals vanish. Discovery increases exactly when $w_k\delta^+>w_j\delta^-$. In particular, equal positive weights make every feasible balancing transfer weakly increase $P_t(q)$ while lowering or preserving the maximum count, strictly when $\delta^+>\delta^-$. Equation~\ref{eq:transfer} describes one transfer, not a campaign.

At fixed candidate and active totals, replacement changes regional discovery but not whole-space discovery.

\subsection{Best Final Counts for the Selected Accounts}
Visa's 8-digit-BIN guidance defines nine-digit account ranges and advises assignment within specified ranges \cite{visa_bin_assignment}. This motivates restricted \emph{domains}, not our bucket sizes. Assume buckets of capacity $M$ in disjoint domains that accounts cannot leave. Let $a_j$ be bucket $j$'s initial count, $S$ the accounts selected at least once, and $r_j$ its initially assigned accounts outside $S$. These unselected accounts retain their numbers.

Selected accounts may finish anywhere in their initial domain, including their original bucket. Retain capacity limits but ignore retirement and intermediate steps. Domain $D$ has $n_D$ buckets and $T_D=\sum_{j\in D}a_j$ accounts.

\paragraph{Proposition (smallest possible maximum count).}
Under these assumptions, the smallest achievable maximum bucket count is
\begin{equation}
  h^*=\max_D\left\{\max_{j\in D}r_j,\ \left\lceil T_D/n_D\right\rceil\right\}.
  \label{eq:endpoint}
\end{equation}
\emph{Proof.} Each bucket retains its unselected accounts, and some bucket in each domain reaches at least the rounded-up domain average. Both terms are therefore lower bounds. Set $h=h^*\leq M$. Domain $D$ has $n_Dh-\sum_{j\in D}r_j$ places available without exceeding $h$ per bucket, enough for its $T_D-\sum_{j\in D}r_j$ selected accounts. Assigning them to these places attains the bound. \hfill$\square$

Monotonicity in active count gives the lower bound $W^*(q)=F(M,h^*,q)$. It does not guarantee that replacement can reach that state.

\paragraph{How many additional accounts are needed?}
Let integer $H$ limit each bucket's final active count. If a domain's rounded-up average exceeds $H$, reassignment within that domain cannot meet the limit. Otherwise, retaining every account in $S$, the minimum number of additional distinct accounts needed in the best-case calculation is
\begin{equation}
  K(H)=\sum_j\max(0,r_j-H).
  \label{eq:additional}
\end{equation}
Each bucket with $r_j>H$ requires $r_j-H$ previously unselected accounts. Selecting that many from each leaves every $r_j\leq H$. Equation~\ref{eq:endpoint} then guarantees feasibility under the simplified assumptions, not completion of the replacement process.

\paragraph{Can all replacements be completed?}
Let $E_D$ count requested replacement events in domain $D$, counting repeated selections separately. If no numbers are initially retired and reuse is forbidden, completion requires
\begin{equation}
  E_D\leq n_DM-T_D\quad\hbox{for every domain }D.
  \label{eq:inventory}
\end{equation}
Each replacement consumes one initially unused number. Equation~\ref{eq:inventory} is necessary, not sufficient: destination restrictions may cause earlier failure. Requested and completed events are reported separately.

\section{Experimental Design}
We compare required replacement, targeted additions, equal-volume random additions, and robustness across densities and initial shapes. The first experiments use 5,000 accounts in 50 buckets. One bucket starts with 100, 400, or 900 accounts; the remainder are balanced. The illustrative cap $H=120$ gives 20\% slack above the mean; neither it nor the 16-choice allocator is an issuer recommendation or required by the exact result.

\paragraph{Assignment and processing.}
Local stays in the current bucket. Group-16 makes 16 draws with replacement from the other nine buckets in its ten-bucket group; 16-choice makes 16 draws with replacement from the other 49 buckets. Choose the eligible draw with fewest active numbers, breaking ties by first appearance. If none has unused numbers, choose uniformly among eligible alternatives; if none exists, stop. Uniform chooses uniformly among eligible alternative buckets globally. All rules sample an unused number uniformly within the destination.

For the supplied-prefix study, a random ordering selects 1,000 required accounts, followed by each bucket's excess unselected accounts above $H=120$. Required replacements occur before additions. Across 30 replications, policies use the same initial assignments, required accounts, and matched destination draws; random additions are selected independently of targeted additions.

\paragraph{Verification and statistics.}
Checks enforce one active number per account, no retired-number reuse, count conservation, and destination capacity. Discovery is recomputed exactly. Paired pointwise 95\% Student-$t$ intervals use 30 test runs \cite{nist_paired_intervals}.

\section{Results}
\paragraph{Feasibility and inventory counterexamples.}
Place 400 of 5,000 accounts in one bucket and balance the remainder. Swapping two accounts elsewhere leaves destinations below the $F(1000,120,10)=72.32\%$ threshold, yet $W(10)=99.41\%$ and $h^*\geq400$: the target remains infeasible. Separately, a bucket with 900 active and 100 unused numbers fails its 101st local replacement, showing that inventory differs from active capacity.

\paragraph{Why the averages can reverse.}
Return to the two-prefix example in Section~\ref{sec:audit}. Let $w$ be the probability of searching the initially denser prefix. Balancing raises discovery exactly when
\begin{equation}
  w<\frac{F(999,250,10)-F(999,100,10)}{F(999,400,10)-F(999,100,10)}\approx0.85485.
  \label{eq:weight-threshold}
\end{equation}
At $w=0.8$ discovery rises by 1.87 points; at $w=0.9$ it falls by 1.54. These are exact sensitivities, not campaign results.

\paragraph{First discovery versus harvesting.}
Full-budget expected yield,
\[
  Y_t(q)=\sum_b w_b qA_t(b)/(M-1),
\]
is linear in counts, unlike $P_t(q)$. For $w=0.8$ and $q=10$, first-hit probability rises from 92.60\% to 94.47\% while yield falls from 3.40 to 2.50. More successful episodes need not mean more harvested accounts.

\subsection{Adding Replacements}
\label{sec:retired-prefix}
Each episode supplies one original card number uniformly from 1,000 randomly selected required accounts, outside the search budget. Search uses its 11-digit prefix, 12-digit prefix, or the whole space. Excluding that original leaves 9,999, 999, or 49,999 candidates, respectively.

Three concentrations, 30 runs, and four rules give 360 campaigns sharing required accounts, order, and initial assignments. Additions use Eq.~\ref{eq:additional}'s unrestricted minimum. Exact probabilities average over all 1,000 supplied numbers and seven budgets. Thirty Local campaigns fail; failure is reported as an outcome, and post-campaign metrics are unavailable for those runs.

With 900 designated accounts, additions to required-only 16-choice reduce supplied-12-digit discovery from 20.32\% to 10.34\% at $q=1$. At $q=10$, it falls from 67.48\% to 66.51\%, a 0.97-point reduction [0.83, 1.12]. Supplied-11-digit discovery falls from 66.82\% to 65.26\%. By contrast, the worst-12-digit benchmark falls by 27.68 points at $q=10$. The additional accounts average \AdditionsMeanFour{} and \CostExtraMean{} per run in the 400- and 900-account cases, respectively. All 60 augmented 16-choice campaigns finish at $h=120$; Group-16 and Uniform do not meet that target.

Thus a policy can buy a large count-control improvement while delivering only a small gain in the attacker-conditioned metric being audited.

\paragraph{Targeting at equal volume.}
We compare targeted additions with the same number of random additions under 16-choice. Each pair retains the same required accounts and initial register and uses matched destination draws. Thirty runs at initial counts 400 and 900 give 120 matched campaigns.

All campaigns finish; targeting meets $H=120$ in all 60 runs, random additions in none. With 900 designated accounts, 12-digit discovery at $q=10$ is 68.26\% for random additions versus 66.51\% for targeting: a paired reduction of \MatchedReduction{} percentage points [\MatchedLow{}, \MatchedHigh{}]. Whole-space discovery is unchanged; targeting slightly increases 12-digit discovery at $q=100$ relative to required-only replacement.

Targeting therefore dominates random replacement on count control and improves discovery at $q=10$, but that ranking is not stable across exposure shapes.

\subsection{Robustness across Shapes and Densities}
\label{sec:robustness}
\begin{table}[t]
\centering\small
\setlength{\tabcolsep}{3pt}
\caption{Targeted minus random supplied-12-digit-prefix discovery at $q=10$ (percentage points; 30 pairs). Positive favors the attacker; brackets give pointwise 95\% intervals. Uniform is shown once.}
\label{tab:robustness}
\begin{tabular}{lrr}
\toprule
Initial shape & Density & Targeted $-$ random (pp)\\
\midrule
Uniform & All & 0.00\\
One hotspot & 2\% & $-7.81$\,[$-8.18, -7.44$]\\
One hotspot & 10\% & $-1.58$\,[$-1.70, -1.46$]\\
One hotspot & 20\% & $0.58$\,[$0.54, 0.61$]\\
Five hotspots & 2\% & $-10.84$\,[$-11.14, -10.54$]\\
Five hotspots & 10\% & $-10.93$\,[$-11.08, -10.78$]\\
Five hotspots & 20\% & $-0.14$\,[$-0.19, -0.09$]\\
Batch-inspired & 2\% & $-0.03$\,[$-0.13, 0.07$]\\
Batch-inspired & 10\% & $0.16$\,[$0.11, 0.21$]\\
Batch-inspired & 20\% & $0.11$\,[$0.08, 0.13$]\\
\bottomrule
\end{tabular}
\end{table}

We cross uniform, one-hotspot, five-hotspot, and batch-inspired counts with densities 2\%, 10\%, and 20\%. Hotspots hold $\min(900,9\bar a)$ or $\min(900,6\bar a)$ accounts; other counts are balanced. Batches fill five-bucket blocks. Locations vary; this models concentration, not the FCA's actual batches \cite[para.~4.7]{fca_tesco_2018}.

The first 20\% of a random ordering are required replacements. From the remainder, targeting selects each bucket's excess above $H=\lceil1.2\bar a\rceil$; the comparator selects the same number of accounts uniformly at random. Each pair starts from the same register and uses matched account orderings and destination draws.

We evaluate supplied 12-digit-prefix discovery at budgets 1, 2, 5, 10, 20, 50, and 100 over 30 paired test runs per setting.

All campaigns finish. Targeting meets $H$ in all 360 test runs; random additions do so only in the 90 uniform runs needing none. At $q=10$, targeting raises supplied-prefix discovery in three cells, with pointwise intervals excluding zero (Table~\ref{tab:robustness}). This recurrence rules out a general security guarantee but is not a fraud-loss or prevalence estimate.

A separate 30-run paired count-level replay uses the same shapes and densities and crosses 4, 8, and 16 choices with cap factors 1.1, 1.2, and 1.4. All nine combinations reverse in three to five of twelve cells; their largest $q=10$ increases are 0.44--0.94 points. Thus reversals persist across allocator parameters; this is not robustness to attacker knowledge or ranking.

\begin{table}[t]
\centering\small
\setlength{\tabcolsep}{3pt}
\caption{Budget sensitivity over 12 supplied-prefix settings. Positive means targeting raises discovery; values at most $10^{-10}$ percentage points count as zero.}
\label{tab:budgets}
\begin{tabular}{lrrrrrrr}
\toprule
Budget $q$ & 1 & 2 & 5 & 10 & 20 & 50 & 100 \\
\midrule
Settings with increases & 2 & 2 & 2 & 3 & 6 & 7 & 7 \\
Largest increase (pp) & 0.026 & 0.056 & 0.117 & 0.579 & 1.129 & 0.425 & 0.970 \\
\bottomrule
\end{tabular}
\end{table}
\paragraph{Budget sensitivity.}
Table~\ref{tab:budgets} summarizes every budget. Reversals extend beyond ten guesses, though some are negligible near saturation. Two settings reverse at $q=1$, where $P_t(1)$ is linear, so exposure weights and allocations also matter.

\section{Discussion and Conclusion}
Count control is operational, not a security guarantee. We do not claim balancing is generally harmful; lower maxima simply do not order attacker-conditioned discovery. Exact analysis proves this non-implication, and matched simulations reproduce it.

Fixed weights and capacities, permanent retirement, recognition noise, and synthetic variation limit inferred magnitude and prevalence, not the exact non-implication. Other exposure, ranking, and yield models may reorder policies; merchant controls may limit attempts \cite{jiang2024enumeration}. No bank data or payment queries were used; issuer studies should estimate weights, recognition error, rankings, range constraints, and reissuance costs.

Audits should record domains, inventory, selection, strategy, budget, and exposure; verify Eq.~\ref{eq:inventory}, completion, and equal-volume comparisons; report count control separately from discovery; treat $h^*\leq H$ as relaxed feasibility; and treat $h\leq H$ as an upper bound on discovery, not as an ordering of $P(q)$ across policies.

\section*{Acknowledgments}
We thank Taro Tsuchiya for helpful early feedback on the paper's framing, terminology, and threat-model presentation.

\clearpage
\bibliographystyle{splncs04}
\begingroup
\sloppy
\hbadness=10000
\bibliography{prefix_count_limits}

@misc{pci_luhn_faq,
  author = {{PCI Security Standards Council}},
  title = {How can {I} validate if a number is a legitimate credit card number?},
  year = {2012},
  howpublished = {FAQ 1137},
  url = {https://www.pcisecuritystandards.org/faqs/1137/},
  note = {Accessed 2026-09-08; distinguishes checksum validity from issuance and activity}
}

@misc{nist_paired_intervals,
  author = {{National Institute of Standards and Technology}},
  title = {Confidence Intervals for Differences Between Means},
  howpublished = {NIST/SEMATECH e-Handbook of Statistical Methods, Section 7.3.1.2},
  url = {https://www.itl.nist.gov/div898/handbook/prc/section3/prc312.htm},
  note = {Paired-observation intervals; accessed 2026-09-05}
}

@article{azar1999balanced,
  author = {Azar, Yossi and Broder, Andrei Z. and Karlin, Anna R. and Upfal, Eli},
  title = {Balanced Allocations},
  journal = {SIAM Journal on Computing},
  volume = {29},
  number = {1},
  pages = {180--200},
  year = {1999},
  doi = {10.1137/S0097539795288490}
}

@inproceedings{bansal2022deletions,
  author = {Bansal, Nikhil and Kuszmaul, William},
  title = {Balanced Allocations: The Heavily Loaded Case with Deletions},
  booktitle = {2022 IEEE 63rd Annual Symposium on Foundations of Computer Science (FOCS)},
  pages = {801--812},
  publisher = {IEEE},
  year = {2022},
  doi = {10.1109/FOCS54457.2022.00081}
}

@misc{stripe_decline_codes,
  author = {{Stripe}},
  title = {Stripe Decline Codes},
  url = {https://docs.stripe.com/declines/codes},
  note = {Technical documentation; accessed 2026-09-05}
}

@article{ali2017distributed,
  author = {Ali, Mohammed Aamir and Arief, Budi and Emms, Martin and van Moorsel, Aad},
  title = {Does the Online Card Payment Landscape Unwittingly Facilitate Fraud?},
  journal = {IEEE Security \& Privacy},
  volume = {15},
  number = {2},
  pages = {78--86},
  year = {2017},
  doi = {10.1109/MSP.2017.27}
}

@article{graves2018too,
  author  = {Graves, James T. and Acquisti, Alessandro and Christin, Nicolas},
  title   = {Should Credit Card Issuers Reissue Cards in Response to a Data Breach?: Uncertainty and Transparency in Metrics for Data Security Policymaking},
  journal = {ACM Transactions on Internet Technology},
  volume  = {18},
  number  = {4},
  pages   = {54:1--54:19},
  year    = {2018},
  doi     = {10.1145/3122983}
}

@misc{fca_tesco_2018,
  author       = {{Financial Conduct Authority}},
  title        = {Final Notice: {Tesco Personal Finance plc}},
  year         = {2018},
  howpublished = {\url{https://www.fca.org.uk/publication/final-notices/tesco-personal-finance-plc-2018.pdf}},
  note         = {Accessed 2026-09-08}
}

@misc{visa_enumeration_2021,
  author       = {{Visa}},
  title        = {{Visa} Guidance to Guard Against Enumeration Attacks and Account Testing Schemes},
  year         = {2021},
  howpublished = {\url{https://usa.visa.com/content/dam/VCOM/global/support-legal/documents/visa-guidance-to-guard-against-enumeration-sept.pdf}},
  note         = {13 September, Article AI11312; accessed 2026-09-08}
}

@techreport{visa_perc_2025,
  author      = {{Visa}},
  title       = {Payment Ecosystem Risk and Control Report},
  institution = {Visa},
  year        = {2025},
  month       = jun,
  url         = {https://corporate.visa.com/content/dam/VCOM/corporate/visa-perspectives/trends-and-insights/documents/visa-perc-report-june-2025.pdf},
  note        = {Accessed 2026-09-22}
}

@misc{visa_vaai_2023,
  author = {{Visa}},
  title = {How {Visa Account Attack Intelligence} Protects Your Business from Enumeration Attacks},
  year = {2023},
  url = {https://corporate.visa.com/en/sites/visa-perspectives/security-trust/2023/how-vaai-protects-your-business-from-enumeration-attacks.html},
  note = {Accessed 2026-09-22}
}

@misc{netcetera_bin_attack_2025,
  author       = {{G+D Netcetera}},
  title        = {Fraud Alert: Inside a Massive {BIN} Attack},
  year         = {2025},
  month        = nov,
  howpublished = {Tech Tribe podcast and video, \url{https://youtu.be/IfHsmJHPOcs}},
  note         = {Published November 18, 2025; accessed 2026-09-22}
}

@misc{jiang2024enumeration,
  author = {Jiang, Bo},
  title = {A Different Type of Card Fraud: Anatomy of a Primary Account Number ({PAN}) Enumeration Attack},
  year = {2024},
  month = nov,
  howpublished = {Privacy Blog, \url{https://www.privacy.com/blog/a-different-type-of-card-fraud-anatomy-of-a-pan-enumeration-attack}},
  note = {Accessed 2026-09-16}
}

@inproceedings{molloy2007dynamic,
  author = {Molloy, Ian and Li, Jiangtao and Li, Ninghui},
  title = {Dynamic Virtual Credit Card Numbers},
  booktitle = {Financial Cryptography and Data Security},
  series = {Lecture Notes in Computer Science},
  volume = {4886},
  pages = {208--223},
  publisher = {Springer},
  year = {2007},
  doi = {10.1007/978-3-540-77366-5_19}
}

@inproceedings{alvim2012gain,
  author = {Alvim, M{\'a}rio S. and Chatzikokolakis, Kostas and Palamidessi, Catuscia and Smith, Geoffrey},
  title = {Measuring Information Leakage Using Generalized Gain Functions},
  booktitle = {IEEE 25th Computer Security Foundations Symposium},
  pages = {265--279},
  publisher = {IEEE},
  year = {2012},
  doi = {10.1109/CSF.2012.26}
}

@article{pendleton2016metrics,
  author  = {Pendleton, Marcus and Garcia-Lebron, Richard and Cho, Jin-Hee and Xu, Shouhuai},
  title   = {A Survey on Systems Security Metrics},
  journal = {ACM Computing Surveys},
  volume  = {49},
  number  = {4},
  pages   = {62:1--62:35},
  year    = {2016},
  doi     = {10.1145/3005714}
}

@inproceedings{bonneau2012guessing,
  author    = {Bonneau, Joseph},
  title     = {The Science of Guessing: Analyzing an Anonymized Corpus of 70 Million Passwords},
  booktitle = {2012 IEEE Symposium on Security and Privacy},
  pages     = {538--552},
  publisher = {IEEE},
  year      = {2012},
  doi       = {10.1109/SP.2012.49}
}

@inproceedings{wang2023silver,
  author    = {Wang, Ding and Shan, Xuan and Dong, Qiying and Shen, Yaosheng and Jia, Chunfu},
  title     = {No Single Silver Bullet: Measuring the Accuracy of Password Strength Meters},
  booktitle = {32nd USENIX Security Symposium (USENIX Security 23)},
  pages     = {947--964},
  publisher = {USENIX Association},
  year      = {2023},
  month     = aug,
  url       = {https://www.usenix.org/conference/usenixsecurity23/presentation/wang-ding-silver-bullet}
}

@misc{visa_bin_assignment,
  author = {{Visa}},
  title = {The 8-Digit {BIN} Expansion Is Coming {April} 2022},
  year = {2022},
  howpublished = {Issuer-processor guidance, \url{https://usa.visa.com/content/dam/VCOM/regional/na/us/partner-with-us/documents/visa-8digitbin-expansion-issuer-processors.pdf}},
  note = {Historical transition guidance; accessed 2026-09-16}
}

@misc{austin_fcu_bin_2025,
  author = {{Austin Federal Credit Union}},
  title = {8-Digit {BIN} Mandate},
  year = {2025},
  howpublished = {Debit-card member notice, \url{https://austinfcu.com/accounts/debit-card/}},
  note = {Describes a twelve-month card-number migration beginning March 2025; accessed 2026-09-22}
}

@techreport{sas2021bin,
  author = {{SAS Institute}},
  title = {{BIN} Attacks: Predicting and Preventing Financial and Reputational Loss},
  institution = {SAS Institute},
  year = {2021},
  month = apr,
  type = {Technical paper},
  url = {https://support.sas.com/content/dam/SAS/support/en/technical-papers/bin-attacks.pdf},
  note = {Last updated April 2021; accessed 2026-09-16}
}

@techreport{allen2025cardsim,
  author = {Allen, Jeffrey S.},
  title = {{CardSim}: A {Bayesian} Simulator for Payment Card Fraud Detection Research},
  institution = {Board of Governors of the Federal Reserve System},
  series = {Finance and Economics Discussion Series},
  number = {2025-017},
  year = {2025},
  month = feb,
  doi = {10.17016/FEDS.2025.017}
}
\endgroup
\end{document}